# Molecular dynamics study of photodissociation of water in crystalline and amorphous ice


Stefan Andersson[a)]

*Leiden Observatory, P.O. Box 9513, 2300 RA Leiden, The Netherlands*

*and Gorlaeus Laboratories, Leiden Institute of Chemistry, Leiden University, P.O. Box 9502, 2300 RA Leiden, The Netherlands*

Ayman Al-Halabi

*Surface Science Research Centre, Department of Chemistry, The University of Liverpool, Liverpool, L69 3BX, United Kingdom*

Geert-Jan Kroes

*Gorlaeus Laboratories, Leiden Institute of Chemistry, Leiden University, P.O. Box 9502, 2300 RA Leiden, The Netherlands*

Ewine F. van Dishoeck

*Leiden Observatory, P.O. Box 9513, 2300 RA Leiden, The Netherlands*



We present results of classical dynamics calculations, performed to study the photodissociation of water in crystalline and amorphous ice surfaces at a surface temperature of 10 K. A modified form of a recently developed potential model for the


---


[a)] Electronic mail: s.andersson@chem.leidenuniv.nl




photodissociation of a water molecule in ice [Andersson *et al.*, Chem. Phys. Lett. **408**, 415 (2005)] is used. Dissociation in the top six monolayers is considered. Desorption of $H_2O$ has a low probability (less than 0.5% yield per absorbed photon) for both types of ice. The final outcome strongly depends on the original position of the photodissociated molecule. For molecules in the first bilayer of crystalline ice and the corresponding layers in amorphous ice, desorption of H atoms dominates. In the second bilayer H atom desorption, trapping of the H and OH fragments in the ice, and recombination of H and OH are of roughly equal importance. Deeper into the ice H atom desorption becomes less important and trapping and recombination dominate. Motion of the photofragments is somewhat more restricted in amorphous ice. The distribution of distances traveled by H atoms in the ice peaks at 6 - 7 Å with a tail going to about 60 Å for both types of ice. The mobility of OH radicals is low within the ice with most probable distances traveled of 2 and 1 Å for crystalline and amorphous ice, respectively. OH is however quite mobile on top of the surface, where it has been found to travel more than 80 Å. Simulated absorption spectra of crystalline ice, amorphous ice, and liquid water are found to be in very good agreement with experiments. The outcomes of photodissociation in crystalline and amorphous ice are overall similar, but with some intriguing differences in details. The probability of H atoms desorbing is 40% higher from amorphous than from crystalline ice and the kinetic energy distribution of the H atoms is on average 30% hotter for amorphous ice. In contrast, the probability of desorption of OH radicals from crystalline ice is much higher than that from amorphous ice.



# I.    INTRODUCTION

The effects of ultraviolet irradiation on water ice are important for the chemistry of both atmospheric[1] and interstellar[2,3] ices. The chemical and physical processes following the absorption of UV photons by condensed phases of water are rather poorly understood, however.

When studying the first UV absorption band of ice, which will be the main focus in this paper, it is of particular interest to understand how the photofragments (for the first excited state, H and OH) behave after photodissociation of a $H_2O$ molecule. The mobility of the species in the ice and at the surface will determine how likely it is for these species to further react with other co-absorbed atoms and molecules. For instance, in interstellar space other molecules such as CO, $CO_2$, and $CH_3OH$ are also present in large amounts in the ices.[2,3] A high flux of UV photons can produce multiple photodissociative events within a narrow range of time and space. In that case photofragments released from different sites in the ice could react with eachother, forming for instance $H_2$ and $H_2O_2$. If the photofragments move large distances before becoming trapped, then further reaction is probable, while recombination of H and OH will naturally lower that probability. Reactions of the photofragments with the $H_2O$ molecules of the ice itself may also be possible. Desorption of the photofragments will affect the chemical composition of the ice and desorption of $H_2O$ molecules will be important for the possibility of 'evaporation' of the ice under UV irradiation. The restructuring of the ice that might follow because of the motion of the photofragments and/or $H_2O$ molecules could lead to phase transformation after a sufficient amount of UV photons has been absorbed by the ice.



Molecular dynamics simulations are particularly well suited to provide insight into these processes on a molecular level.

Water ice is abundant in the cold, dense inner regions of interstellar clouds where it has mostly been observed in amorphous phases even though crystalline ice may also be present.[2,3] The ice is frozen onto silicate and/or carbonaceous cores of particles ('dust grains') with a typical size of 0.1 μm. A typical temperature in these regions is 10 K. The relative importance of ultraviolet radiation and thermal processes on the chemistry of the grains is under debate. Ultraviolet photons cannot penetrate into the inner regions of clouds so they are supposedly created by interactions between cosmic rays and $H_2$. The energies of the photons emitted cover the first absorption band of ice as well as higher energies (see Ref. 4 and references therein). However, the UV photon flux is predicted to be very low (about $10^3$ photons $cm^{-2}$ $s^{-1}$), which would give about one incident photon per month on a grain. Thus, the photodissociation dynamics, which occurs on a picosecond timescale, is completed before the next photon arrives. Chemical evolution occurs on a time scale of $10^3$-$10^7$ years, so certain photoinduced processes with high probabilities upon absorption of UV photons may still be important.

The ultraviolet absorption spectra of liquid water[5-8] and several phases of water ice[5,8-10] have been well characterized experimentally. The lowest absorption band has its peak at 8.2 eV in liquid water[6] and at about 8.6-8.7 eV in crystalline as well as amorphous ice phases.[5,8,10] These peaks are blue-shifted compared to gas-phase $H_2O$, which has an absorption peak at 7.4 eV.[11] The threshold energy, below which absorption becomes



insignificant compared to the peak, is found at about 7.6 eV for both crystalline and amorphous ice[10], while the threshold for liquid water lies a few tenths of an eV lower in energy.[5-8] However, for liquid water there is very weak absorption down to about 6 eV, the so-called Urbach tail,[5,8,12,13] the origin of which remains unclear. Similar, but less pronounced, tails have also been found in the spectra of ices.[8,14]

In crystalline and amorphous ices a number of products have been observed after UV irradiation. Ghormley and Hochanadel[14] observed H, OH, and $H_2O_2$ after flash photolysis of crystalline ice at 263 K. Gerakines et al.[15] exposed amorphous ice ($T = 10$ K) to UV light with energies corresponding mainly to the first and second absorption bands and observed OH, $HO_2$, and $H_2O_2$ as products in the condensed phase through infrared spectroscopy. Their experimental technique did not allow the detection of H atoms, however. After irradiation of amorphous ice surfaces at $T = 35$-100 K by Lyman-$\alpha$ photons (10.2 eV) Westley et al.[16,17] observed desorption of $H_2O$ molecules, but only after relatively large UV doses ($10^{18}$ photons cm$^{-2}$). The desorption probability was found to increase for higher UV doses until reaching plateau values of about $10^{-3}$-$10^{-2}$ desorbing $H_2O$ per incident photon. Desorption of $H_2O$ was also found to increase with increasing temperatures. At the lower temperatures no desorbing $H_2O$ could be found in the limit of single-photon absorption. Other species, such as $H_2$ and $O_2$, were found to desorb and upon heating yet different species, possibly OH and $H_2O_2$, were detected.[16] Watanabe et al.[18] irradiated amorphous $D_2O$ ice ($T = 12$ K) with 7.2 eV and 9.8 eV light. At the higher energy $D_2$ was readily formed, but at the lower energy very little $D_2$ was detected. It was inferred that the $D_2$ was mostly formed by recombination of trapped D atoms and



occurred more readily upon heating of the ice. Desorption of $D_2O$ was found to be negligible. Baggott et al.[19] studied the effect of 21 eV and 41 eV photons on a $D_2O$ ice surface at $T = 80$ K. At both energies $D^+$ was found to desorb and at the higher energy also $D_3O^+$ was detected.

In the case of liquid water there is an ongoing debate on the nature of the 'hydrated electron' that is formed after UV irradiation.[20-22] Recent quantum chemistry calculations suggest that this actually could be a transient $H_3O$ species,[23-25] but conclusive evidence has yet to be found. Other products that have been detected after UV irradiation of liquid water include H, OH, and $H_3O^+$.

A few studies have also dealt with excitation energies below the threshold, i.e. in the low-energy tail of the spectrum. Quickenden et al. (see Refs. 26 and 27 and references therein) have made extensive studies of the products detected in ice after irradiation of 5.6 eV and 4.8 eV photons. Through studying the luminescence they were able to identify products such as OH and $O_2$ but there were also bands of uncertain origin. Langford et al.[27] assigned luminescence at 420 nm to OH, while Mathers et al.[28] proposed that it was due to a $H_3O$ species. Yabushita et al.[29] studied the production of H atoms after irradiation of amorphous ice at 90-140 K by 6.4 eV UV light. They were able to detect desorbing H atoms with a kinetic energy distribution with two peaks, at 0.39 eV and at 0.02 eV, respectively. With the aid of *ab initio* (CASSCF) calculations, they attributed their observation to photodissociation of $H_2O$ molecules at the surface.



In a study on irradiation of thin ice films (up to 10 monolayers) on graphite by UV photons with energies lower than 5.6 eV, Chakarov and Kasemo[30] observed crystallization of the initially amorphous ice. The mechanism was proposed to involve electrons released into the ice after excitation of the underlying graphite substrate rather than absorption of photons within the ice. Leto and Baratta[31] found evidence for amorphization of crystalline cubic ice at 16 K after irradiation by Ly-$\alpha$ photons.

A number of theoretical studies have dealt with calculating the electronic excitation energies and absorption spectra of water clusters,[32-34] liquid water,[35-37] and ice[38] with more or less sophisticated methods. However, only our recent paper[39] has explicitly dealt with the dynamics following absorption of UV photons. In that paper we presented results for molecular dynamics simulations of the photodissociation of $H_2O$ molecules in crystalline ice at $T = 10$ K.[39] In this paper, we have extended that model to both crystalline and amorphous ice to investigate the influence of surface morphology on the photodissociation dynamics. Our interest in photodissociation of water in ice comes from an interest in reactions of the photofragments with co-adsorbed species. Therefore our study focuses on the dynamics of photodissociation in the uppermost layers of the ice.

In Section II details of the computational methods are presented, regarding the setup of ice surfaces, the potentials used, and the initialization, propagation, and termination of the photodissociation calculations. Section III contains the results that have been obtained on trapping, recombination, desorption, and mobility of photofragments. Finally, in Section



IV the most important results are summarized and suggestions for future work are presented.

## II.    COMPUTATIONAL DETAILS

### A. Setup of ice surfaces

Both crystalline and amorphous ice surfaces have been constructed. The crystalline ice model describes normal hexagonal ice ($I_h$) and consists of a slab of eight bilayers (sixteen monolayers) of 60 (30) $H_2O$ molecules each. These molecules are treated as rigid rotors. The molecules in the top six bilayers are allowed to move, while the two bilayers in the bottom are kept fixed to simulate bulk ice. To simulate a (0001) basal plane infinite ice surface, periodic boundary conditions are applied in the $x$- and $y$-directions, $z$ being the surface normal (see Fig. 1). The dimensions of the unit cell are 22.4 Å, 23.5 Å, and 29.3 Å in $x$, $y$, and $z$, respectively. The zero of the $z$ coordinate is set at the bottom of the ice slab. The initial ice configuration obeys the ice rules[40] and has a zero dipole moment. This surface was equilibrated using a Molecular Dynamics (MD) code, employing a leapfrog algorithm, as has been described in Ref. 41, with the TIP4P potential,[42] which is described in Section II.B. The equilibration was run at $T = 10$ K for 100 ps.

The amorphous ice model was constructed in a similar fashion to that used by Al-Halabi et al.[43] The surface was initially set up as crystalline ice, but instead of equilibrating for 100 ps at $T = 10$ K, the ice was equilibrated for 5 ps at $T = 10$ K, after which the ice was heated to 300 K during an interval of 20 ps using the computational analog of a



thermostat. This liquid phase was equilibrated for 110 ps and then it was cooled rapidly (during 20 ps) to 10 K. Thereafter it was once again equilibrated for 120 ps.

In Figs. 1 and 2 top and side views of the six top monolayers, containing 180 $H_2O$ molecules, are shown for the crystalline and amorphous ice models obtained with the above procedure. As can be seen in Fig. 1 the crystalline ice exhibits hexagonal 'shafts' that run along the $z$-direction, and the molecules are ordered in bilayers that are separated by about 3.5 Å. The amorphous ice shown in Fig. 2 has no such regular behavior. The ice does exhibit some irregular ring structures but the binding properties differ from that of crystalline ice. In the top monolayer of crystalline ice the molecules are three-coordinated but further into the ice all molecules are four-coordinated. Amorphous ice has a more irregular bonding structure and at the surface there are a few molecules that are only two-coordinated (see also Ref. 43 for a detailed discussion). The most marked difference between the ices is to be found in the top monolayers. For instance, the top monolayer is much less dense in amorphous ice, where it stretches from a value of $z$ of 26.3 Å to 29.2 Å, with the positions referring to the centers of mass of the top 30 molecules. The corresponding monolayer in crystalline ice is found between 28.2 Å and 28.9 Å.

## B. Potentials

The ice surfaces are constructed using the TIP4P potential[42] with all molecules kept rigid. This potential consists of O-O Lennard-Jones (LJ) interactions and electrostatic interactions based on charges situated on the H atoms and on an additional charge site M close to the O atom (M: -1.04$e$; H: 0.52$e$). To calculate the photodissociation of one of



the H$_2$O molecules (Section II.C) one of the rigid molecules is exchanged with one that is fully flexible, meaning that its atoms are allowed to move without any dynamical constraints. To avoid the computational difficulties involved in incorporating the massless charge site M into a framework of freely moving atoms, a model with only atom-centered interactions is used for describing the interaction of the H$_2$O molecule to be photodissociated with other H$_2$O molecules. The choice was made to use the TIP3P model[42] for the intermolecular interactions of the flexible H$_2$O molecule, in its electronic ground state, with the TIP4P molecules. The TIP3P potential is similar to TIP4P with the difference that it only has atom-centered charges (O: -0.834$e$; H: 0.417$e$) and slightly different O-O LJ parameters. The ground state TIP3P interaction is used in the calculation of the absorption spectrum (to compute excitation energies while taking into account interactions with surrounding H$_2$O molecules) and in the dynamics (whenever the possibility of recombination arises, see below).

The intramolecular potential for the ground and first excited states is taken from the potential energy surfaces (PESs) constructed by Dobbyn and Knowles (DK)[44] for gas-phase H$_2$O. In order to simulate the intermolecular interactions of the H$_2$O molecule in its first excited state, atomic charges (O: 0.4$e$; H: -0.2$e$), which in our previous study[39] were derived from a calculated dipole moment of the first excited state of gas-phase H$_2$O,[32,45] were put on the O and H atoms. This will henceforth be referred to as the 'old potential'. This together with the standard TIP3P LJ potential gave an absorption spectrum that was shifted too much to the blue (see Section III.A). To bring the excitation energies into agreement with the experimentally measured absorption peak of crystalline ice the



charges have been adjusted to new values (O: -0.2$e$; H: 0.1$e$). This will be called the 'new potential'.

The total potential energy for the ice in the photodissociation calculations can thus be written as:

$$V_{tot} = V_{ice} + V_{H_2O^*-ice} + V_{H_2O^*} \qquad (1)$$

$V_{ice}$ is the intermolecular interaction of the $H_2O$ molecules excluding the dissociating molecule. This is described using the TIP4P model referred to above. The second term is the intermolecular interaction of the molecule, which is to be dissociated, is dissociating, or has recombined, with the rest of the ice using the (modified) TIP3P model and the last term is the intramolecular $H_2O$ potential, described using the appropriate DK PES.

Upon photoexcitation, the $H_2O^*$-ice interaction is first given by a TIP3P-type potential with the charges on O and H as discussed previously. When the molecule dissociates, the TIP3P-type intermolecular potential at some point has to be switched into separate potentials for the H-$H_2O$ and OH-$H_2O$ interactions.

The H-$H_2O$ potential has been calculated as a reparameterization of the very accurate YZCL2 gas-phase $H_3O$ PES,[46] which was interpolated using MRCI and CCSD(T) energy points. The potential form of the present fit consists of H-H and H-O dispersion terms, H-H and H-O repulsive interaction terms, and a H-O Morse potential reflecting the 'partially bonding' character of the H-$H_2O$ interaction. This potential is to the best of our knowledge the most reliable H-$H_2O$ pair potential available, as discussed in a document



of supporting material available through the journal's Electronic Physics Auxiliary

Publication Service (EPAPS).[47]

The OH-$H_2O$ potential has a similar form and consists of dispersion, repulsion and

electrostatic contributions. It was constructed in a simpler way than the H-$H_2O$ potential.

Similarly to the development of the Kroes-Clary HCl-$H_2O$ potential,[41] the parameters for

the repulsive interactions were derived from Hartree-Fock calculations of $H_2O$-$H_2O$

interactions.[48] The O-O dispersion interaction was taken to be the same as in the TIP3P

Lennard-Jones potential and a damping was applied as devised by Ahlrichs et al.[49] For

the electrostatic interaction O and H atomic charges were fitted as a function of O-H

distance to OH dipole moments calculated at the MRCI level.[50] This simple procedure

gives very good agreement with energy points calculated at the CCSD(T) level for long-

range interactions, while at shorter range the attractive interaction is somewhat

underestimated (see the supporting information[47] for details).

To smoothly connect the different parts of the potentials, i.e. between where the

dissociating $H_2O$ is still intact to where it is dissociated into H and OH fragments, a

number of switching functions have been devised. The switching functions allow for

instance the polar $H_2O$ molecule to be switched into a neutral H atom and an OH radical

with a charge distribution different from TIP3P. The switching functions are presented in

the supporting material.[47]



For the dissociating $H_2O$ molecule the intramolecular PES is initially the first excited state PES. When the molecule is dissociated and the first excited state has become near degenerate to the ground state (both states correlate asymptotically to $H(^2S)$ + $OH(X^2\Pi)$), a smooth switch is made to the ground state PES. This is made on the assumption that an internal conversion will be efficient through large mixing of the electronic states, possibly induced by interaction with the surrounding molecules. The potential in the switching region is written as a linear combination of the ground and excited-state PESs. The switch is made when the OH distance is between 3.0 Å and 3.5 Å. These distances have been chosen because this is where the excited-state and ground-state PESs become near degenerate and the switch can be made without introducing troublesome kinks in the potential, the difference in energy between the PESs being smaller than 0.05 eV. If the dissociation of the OH bond should be reversed before reaching 3.5 Å, the system will go back to the excited state PES if the bond becomes shorter than 3 Å. When the OH bond becomes longer than 3.5 Å the system will remain on the ground state PES. This allows for recombination of H and OH. The recombination probability obtained through this procedure should be an upper bound to the 'real' probability due to the fact that the H and OH can also come together on other PESs than the ground state PES (analogous to the gas-phase system[51]), and the possibility that the coupling leading to internal conversion to the ground state is weaker than implied in the present formalism.

The $H_2O$-$H_2O$, OH-$H_2O$, and H-$H_2O$ interactions are all set to zero at distances >10 Å through cutoff functions[41] to avoid interactions with multiple images of the $H_2O$ molecules. The dissociating molecule and the H and OH fragments do not interact with



the $H_2O$ molecules in the frozen layers at the bottom of the slab. The rationale behind this is that these molecules do not respond dynamically to the interaction with the sometimes very energetic photofragments, making the dynamics 'flawed'. Note that the intramolecular $H_2O$ potential is not switched off and this can therefore lead to artificial recombination of H (OH) with a periodic image with OH (H), the probability of which was however observed to be very small (less than 1%).

Regarding the $H_2O$(excited state)-$H_2O$(ground state) potential, calculations have shown that the main reason for the blueshift of the $H_2O$ optical absorption spectrum in the condensed phase is due to an excited-state electronic wave function that extends much further than that of the corresponding gas-phase molecule.[38] This effect appears due to interaction with the neighboring molecule. The electron-hole pair binding energy (between the excited electron and the hole left at the oxygen) is decreased by about 2 eV upon condensation and the average electron-hole distance changes from 2.3 Å to 4.0 Å. These findings imply that (i) there may be significant exchange-repulsion interactions of the excited state molecule with its neighbors, (ii) the intramolecular excited state $H_2O$ potential will be somewhat different from the gas phase (see Ref. 34), and (iii) the excited state molecule will have a very large polarizability. None of these effects are explicitly treated by our potential, but they are included in an *average* way (see Section III.A). The polarizability is in fact accounted for by changing the dipole moment of the excited-state $H_2O$ molecule such that its interaction with the surrounding $H_2O$ molecules becomes less repulsive. Of course the use of static charges cannot exactly mimic the induced dipole



moments for all different molecules in the ice, but the adjustment of the average excited

state dipole moment should represent a good first approximation.

It should also be mentioned that in the case of strong H-$H_2O$ interactions, as in the

collisions with $H_2O$ molecules of a highly energetic H atom released from

photodissociation of $H_2O$, the collision complex and its solvation shell should ideally be

treated as a (metastable) $H_3O$ complex stabilized by the surrounding $H_2O$ molecules. The

interaction of $H_3O$ with neighboring $H_2O$ molecules has been shown to be similar to that

of the excited-state $H_2O$ intermolecular interactions, with a very diffuse electron

distribution and a less destabilized $H_3O$ complex (see Ref. 25 and references therein).

Sobolewski and Domcke[25] calculated the $H_3O$ species in a small water cluster to be

unstable by 0.6 eV, but to have a 0.3 eV barrier to dissociation, compared to 0.8 eV and

0.1 eV for gas-phase $H_3O$.[24] They argue that their results can also explain liquid-phase

experiments, but exactly how well this model compares to the actual condensed-phase

case is not entirely clear. These subtle effects are not described by our potentials and it is

unclear what the effects on the dynamics would be of their inclusion. However, modeling

the H-ice interaction based on accurate H-$H_2O$ interactions for isolated $H_2O$ may be

viewed as the first important step towards describing $H_3O$ in the condensed phase.

## C. Photodissociation calculations

To study the photodissociation of individual $H_2O$ molecules, the crystalline and

amorphous ice surfaces described in Section II.A were used. For each of the top six

monolayers 12 $H_2O$ molecules were chosen to be photodissociated in crystalline ice and



13 molecules in amorphous ice. The dissociating molecules in crystalline ice were picked to give an even distribution of the four distinct orientations of the molecules. The initial state of the ice surface was always taken to be the final time step from the setup run. For each dissociating molecule 200 classical trajectories were run. Therefore, in total 14400 trajectories were run for crystalline ice and 15600 for amorphous ice. The results from gas-phase photodissociation calculations were initialized in the same way but here 1000 initial conditions were sampled and the trajectories integrated using the DK excited state PES.

To initialize the photodissociation trajectories, a semiclassical (Wigner) phase-space distribution fitted to the ground-state vibrational wave function of gas-phase $H_2O$[52] has been used. Initial coordinates and momenta of the atoms in the dissociating $H_2O$ molecule were sampled by a Monte Carlo procedure. Thus the initial conditions are approximately quantum mechanical even though the dynamics is treated fully classically. This approach has been successful for several systems in gas-phase photodissociation, especially for purely repulsive states like the A-state of $H_2O$ considered here.[53] The use of the gas-phase vibrational wave function as starting point for $H_2O$ in ice is assumed to be a minor approximation. Once the initial conditions were set, a Franck-Condon excitation was made and the system was put on the excited state PES as described in Section II.B. To incorporate this molecule into the ice framework the geometry of the molecule chosen to be photodissociated was fixed by exchanging the initial TIP4P geometry with the $H_2O$ geometry generated from the sampling procedure. This was done



by making the molecular plane, the centers of mass, and the bisectors of the HOH angles coincide.

Thereafter the integration of the trajectory was started. The trajectories were run for a maximum time of 20 ps with a time step of 0.02 fs. The trajectory was terminated if all three of the atoms met one of the following conditions: (i) the atom was more than 11 Å above the surface, (ii) the atom was at a negative $z$ (below the ice slab), and (iii) the kinetic energy of the atom was smaller than the absolute value of the intermolecular potential it experienced (given the potential was negative). Note that the O and H atoms in OH and $H_2O$ were treated individually in this scheme. The potential evaluated in (iii) did not include the intramolecular part of the potential (from the DK PES). Therefore, also a highly vibrationally excited OH or $H_2O$ could be considered to be trapped at the outer turning points, i.e. when its intramolecular kinetic energy is low.

The absorption spectra presented in Section III.A have been calculated using the classical approximation.[53] Excitation energies were calculated as the difference of the total potential energy of the ice between the excited state and the ground state for all the initial geometries used in the trajectory calculations. The liquid water spectrum was calculated using the liquid water model used to make the amorphous ice (Section II.A) and calculating initial coordinates as above for molecules corresponding to monolayers 5 and 6. The excitation energies were binned into 0.05 eV wide energy intervals and weighted with the square of the gas-phase transition dipole moment, taken from a coordinate-



dependent function.[54] Calculations on the transition dipole moment of $H_2O$ in the condensed phase[37] suggest this to be a reasonable approximation.

## III.    RESULTS AND DISCUSSION

### A.  Spectrum

In our previous paper[39] it was shown that the absorption spectra of the top bilayers of crystalline ice converged already in the second and third bilayers. Therefore, the third bilayer was used to represent the bulk. However, the adopted model with charges for the excited-state molecule taken from a calculated gas-phase dipole moment gave a blue shift of the crystalline ice spectrum compared to the gas phase of about 2 eV. Since the experimental blue shift is about 1 eV,[10] the dipole moment of the excited state molecule was adjusted (see Section II.B) to make the peak of the crystalline ice spectrum (calculated for the third bilayer) coincide with the experimental value (Fig. 3a). In doing so, also the low-energy threshold of the spectrum was well reproduced. Even more interestingly, the calculated spectra for amorphous ice and liquid water (Fig. 3b) have peaks at 8.6 eV and 8.2 eV, respectively, in excellent agreement with experiments.[7,10] The surprising ability of the current model to reproduce the experimental peaks suggests that the potentials give a good description of the excitation energies in an *average* sense, as discussed in Section II.B. This means that even though potentially important contributions to the potential are not treated explicitly, they are included in an average way. Because the excitation energies now agree more closely with experimental values than in our previous study, we also expect that the dynamics will be better described.



**B. Product fractions and energy distributions: old vs. new potential**

With the introduction of the new charges in the potential the results for the product fractions and energy distributions for crystalline ice changed somewhat compared with our previous paper. However, the changes are not dramatic, as can be seen in Fig. S5.[47] The final outcomes stayed almost exactly the same and the only statistically significant change is that the recombination probability increased slightly. This can be understood through the fact that, with the new potential model, the H atoms are released with a lower kinetic energy and therefore will escape the 'cage' to a lesser extent. The lowered kinetic energy is demonstrated in Fig. S6[47] where a comparison is made between the kinetic energy distributions of the desorbing H atoms originating from the first bilayer for the old and new potential models, respectively. Clearly, going to the new model the peak shifts to lower energies by about 1.4 eV, which is actually more than the 1 eV shift in the peak of the absorption spectrum. As stated before, the dynamics results obtained with the new potential model are believed to be the most reliable and from now on we will only present results for this model.

**C. Basic outcomes**

The probabilities of the most important outcomes of the photodissociation event in crystalline and amorphous ice are plotted in Fig. 4 as function of monolayer. The categories in the figure are the following: 'H desorbs' means that, following photodissociation, an H atom desorbs from the surface while the OH fragment remains trapped, 'H+OH trapped' is the event where both the H and OH become trapped at different locations, 'HOH recombination' is where H and OH recombine and become



trapped as a $H_2O$ molecule, and 'Other' is a collection of events, some of which are artefacts of the model. For crystalline ice all 'Other' events in monolayers 3 through 6 are artefacts, whereas this is the case for monolayers 4 through 6 in amorphous ice. The 'Other' events are discussed below.

Overall, what is most striking is that the outcomes and trends are very similar for amorphous and crystalline ices, although there are some subtle differences in some of the details (Fig. 4). H atom desorption with OH remaining trapped dominates in the top monolayers and as expected the probability of desorption drops when moving deeper into the ice. The desorption probability is clearly higher for the first three monolayers of amorphous ice (89%, 75%, 55%) than for the corresponding layers in crystalline ice (71%, 57%, 27%), but for monolayers 4 to 6 the differences become less pronounced (30%, 22%, 12% for amorphous ice; 22%, 15%, 11% for crystalline ice). Not surprisingly, the crystalline ice shows a decrease in desorption probability in steps of two monolayers, reflecting its bilayer structure, while amorphous ice shows a nearly linear depth dependence with a change of behavior between the third and fifth monolayers. The substantial desorption of hydrogen atoms agrees well with the findings of Westley et al.[17] in their experiments on Lyman-α photon (10.2 eV) irradiation of ice where the production of hydrogen-deficient radicals such as $HO_2$ and $H_2O_2$ pointed to loss of H atoms. However, their excitation energy is somewhat different from the ones in this work, and probably $H_2O$ molecules are excited to higher electronic states in this experiment. The experiments by Gerakines et al.[15] also showed these species being formed from UV irradiation of ice, using radiation which includes the wavelengths studied here.



The probability of dissociation followed by trapping of both H and OH increases with increasing depth. For the first monolayer in crystalline ice the probability of trapping is about 14%, while it is only 2% for the first monolayer in amorphous ice. In crystalline ice it is predominantly the molecules with both hydrogens involved in bonding that contribute to this trapping, with a probability of over 20%, while dissociation of the molecules with dangling H atoms only give trapping in 5% of the cases. In the fifth and sixth monolayers the trapping probability approaches about 50 % and 40 % for crystalline and amorphous ice, respectively. It is interesting to note that the sum of the probabilities of trapping and recombination of H and OH is 80% for both types of ices, which means that recombination is less probable than trapping in crystalline ice (30%), but that the two events are equally probable in amorphous ice (40%). The reasons for this difference will be discussed in Section III.G. The average time from photodissociation until H and OH are trapped (accomodated) is about 1 ps for both types of ice.

 Recombination with subsequent trapping of the recombined molecule has very low probability in the two top monolayers of both crystalline and amorphous ice (less than 1 %) as can be seen in Fig. 5 (where, in addition, the 'Other' category has been split up into four different categories). As seen in the same figure the recombination probability increases substantially in both types of ice when going to the second monolayer (8% for crystalline and 5% for amorphous ice).



Another interesting observation is that desorption of OH is much more probable from crystalline ice than from amorphous ice. In fact 9% of the trajectories originating in the first crystalline monolayer lead to OH desorbing, while this only happens about 2 % of the times in the first amorphous monolayer. Desorption of $H_2O$ will be discussed in detail in Section III.D.

The category called 'Rest' in Fig. 5 almost exclusively consists of artefacts of the model used. These artefacts include H leaving through the 'bottom' of the surface, the trajectory reaching the maximum time $t = 20$ ps without any of the termination criteria from Section II.C being met, or H recombining with OH in a periodic image of the cell. In two trajectories (out of 14400) for crystalline ice $O(^1D)$ and $H_2$ are formed from photodissociation after recombination of H and OH and subsequent redissociation. The oxygen atom in these cases remains trapped and $H_2$ in one case desorbs and in the other case stays trapped in the ice. These outcomes are not artefacts since the $O+H_2$ product channel is well treated by the DK PES.

**D.    $H_2O$ desorption**

As can be seen in Fig. 5 desorption of the recombined $H_2O$ following photoexcitation in monolayers 1 and 2 (and also 3 in amorphous ice) is about equally likely for both types of ice (0.2%). The probability of desorption of the surrounding $H_2O$ molecules is of similar magnitude to the desorption of recombined $H_2O$. Desorption of molecules from monolayers 1 and 2 has been observed in 0.2% of the trajectories for crystalline ice, while it only happens in 0.04% of the cases for amorphous ice. Photodissociation of



molecules in monolayers 2 to 6 in crystalline ice and 1 to 4 in amorphous ice has been found to lead to surface $H_2O$ molecules desorbing. The mechanism of desorption is that a H atom with large kinetic energy transfers part of its momentum to a $H_2O$ molecule in the correct direction. If the molecule hit is at the surface this impulse might 'kick' that molecule off the surface or otherwise the impulse is transferred to the surface by consecutive $H_2O$-$H_2O$ collisions. The higher desorption probability for crystalline ice probably arises from the momentum being more efficiently transferred in a specific direction, such as towards the surface, through the ordered hydrogen-bonded network than is possible in the more disordered amorphous ice. Comparison with the photodesorption experiments by Westley et al.[17] is difficult since their excitation energies were different. They also did not see any desorption in the limit of single-photon absorption in the ice.

An interesting aspect is the efficiency of kinetic energy transfer from H to $H_2O$. The ratio of the energy transferred to the 'collision energy' can in a simple kinematical hard-sphere model[55,56] be written as:

$$r = \frac{\Delta E}{E_i} = \frac{4\alpha}{(1+\alpha)^2} \qquad (2)$$

Here $\Delta E$ is the transferred energy, $E_i$ is the energy with which H hits $H_2O$, and $\alpha$ is the mass ratio of the colliding particles. Since the potential energy experienced by the H atom between collisions can vary by considerable amounts, a special convention is used. The initial energy is equal to the kinetic energy at the minimum H atom potential energy before the collision, if that potential energy is lower than or equal to the minimum potential energy directly after the collision. Should the minimum potential energy after



the collision, $V_{\min,f}$, be lower than the previous minimum potential energy, $V_{\min,i}$, then the difference between the two potential energies is added to the initial kinetic energy, $E_{k,i}$:

$$E_i = E_{k,i} + V_{\min,i} - V_{\min,f} \qquad (3)$$

For energy transfer between a H atom and the center-of-mass motion of a $H_2O$ molecule, $r$ is 0.2.

The maximum energy transferred that is within the calculated $r = 0.2$ has been found to be 0.66 eV. The smallest binding energies of $H_2O$ molecules in the first monolayer of crystalline ice are 0.8 eV. In amorphous ice the lowest binding energy is 0.6 eV and in both types of ices the average binding energy is 0.9 eV in the first monolayer. Of course, strictly speaking, the hard-sphere mode employed here is not rigorously applicable to our model, but the results suggest that one reason for the rather low 'indirect' desorption probability is that most often insufficient translational energy can be transferred to a surface $H_2O$ molecule due to the mass mismatch between H and $H_2O$.

### E.      H atom distance and kinetic energy distributions

Knowledge of how far the photofragments can move through the ice can help us assess to what extent these fragments may undergo subsequent reactions with co-adsorbed molecules upon release into the ice. In the following sections the results have been categorized in terms of bilayers. That means that monolayers 1 and 2 constitute bilayer 1, monolayers 3 and 4 make up bilayer 2, and monolayers 5 and 6 bilayer 3. Of course, the concept of bilayer is not strictly applicable to amorphous ice but to make the proper comparison to crystalline ice the above definition was chosen. In Fig. 6 the distribution of



distances traveled by the H atoms from their original locations are shown for each bilayer of crystalline and amorphous ice. The crystalline ice distance distribution shows more structure than the corresponding one for amorphous ice. This is because of the more regular distribution of favorable binding sites in crystalline ice. The distribution in amorphous ice is shifted somewhat towards shorter distances compared to crystalline ice. The average distances traveled by trapped H atoms in crystalline ice are 10 Å, 10 Å, and 9 Å for bilayers 1, 2, and 3, respectively, and the corresponding numbers are 9 Å, 8 Å, and 8 Å for amorphous ice. The maximum distance the H atom travels is over 60 Å for both types of ice.

The differences in average mobility could at first glance be attributed only to differences in the structure of the ices, with the crystalline ice having 'shafts' that might facilitate the motion of particles through the ice. The explanation however seems to be more involved. Analysis of the loss of kinetic energy of the H atoms during their motion through the ice, shows that the energy transfer to the ice phonon modes is more efficient in amorphous ice than in crystalline ice. The trapped H atoms have on average collided 27 times with $H_2O$ molecules in crystalline ice, a collision being defined to occur if a close encounter gives an interaction energy of H with $H_2O$ larger than zero (zero being the asymptotic energy for $H$-$H_2O$ interactions). At low positive interaction energies (<0.1 eV) the H atom may not lose but actually gain kinetic energy between collisions since the attractive (negative) interactions start to be of similar magnitude to the repulsive interactions. A hard collision is therefore defined to occur if the interaction is 0.1 eV or higher. In crystalline ice the trapped H atoms on average experience 17 hard collisions. The corresponding numbers of



collisions and hard collisions in amorphous ice are 23 and 13, respectively. The average energy lost per hard collision for the trapped H atoms is 0.11 eV in crystalline ice and 0.17 eV in amorphous ice, i.e. for all hard collisions 1.8 eV for crystalline ice and 2.2 eV for amorphous ice. The difference in energy transfer efficiency could be a consequence of the different phonon spectra for the two types of ice. For instance, amorphous ices have an onset of the librational band (about 50-120 meV) shifted to lower energies compared to ice $I_h$,[57] and the lower phonon frequencies could make it easier to transfer energy to these phonons.

One important experimental observable is the kinetic energy distribution of desorbing atoms and molecules. Kinetic energy distributions of the desorbing H atoms have been monitored and are presented in Fig. 7. It is seen that the H atom kinetic energy distribution is for each bilayer hotter for amorphous ice than for crystalline ice. The average kinetic energies for the three bilayers are (from 1 to 3) 1.9, 1.4, and 1.1 eV for amorphous ice and 1.6, 1.0, and 0.8 eV for crystalline ice. This seems to be in contradiction with the finding that H atom kinetic energy loss is more efficient in amorphous ice. Here the structure of the ice seems to become more important instead. The reason for the hotter kinetic energy distribution in amorphous ice is that fewer H-$H_2O$ collisions occur on the way to the surface in amorphous ice than in crystalline ice. The difference is largest for photodissociation in monolayers 3 and 4, where the desorbing H atom in crystalline ice on average has experienced 5 and 6 hard collisions, respectively. The corresponding numbers of collisions in amorphous ice are 3 and 4. As was mentioned in Section II.A the first monolayer is much less dense in amorphous ice



than in crystalline ice, making H atom motion less restricted. This also helps to explain the high H atom desorption probability for the three top monolayers in amorphous ice.

## F. OH distance and vibrational energy distributions

The motion of the OH radicals is naturally more restricted than that of the H atoms as can be seen by comparing Fig. 8, which shows the distribution of the displacements of the O atoms in OH from their original position, to Fig. 6. The results for the first bilayer show an interesting difference between crystalline and amorphous ice (Fig. 8). Whereas the OH distance distribution for amorphous ice peaks at 1 Å, the crystalline distribution has two peaks, at 0.4 Å and 1.8 Å. The explanation for this behavior can be understood by considering Fig. 9. Here the distance distribution is plotted for the separate *monolayers* that make up bilayer 1. The difference between the amorphous monolayers is small, while the difference between monolayers 1 and 2 in crystalline ice is rather large. The peak at short distances is clearly due to molecules in monolayer 1 and comes from those three-coordinated molecules with dangling H atoms (50% of monolayer 1). These H atoms can desorb directly into the gas phase with the OH basically staying at its original location. Since the desorbing atom was not part of any hydrogen bond, this means that OH stays in an energetically favorable position with the original hydrogen bonds intact. In cases where the other, hydrogen bonded, H atom is the one that is released, the remaining OH is only two-coordinated and in most cases, because of the weak binding, it desorbs or travels large distances (up to tens of Å) over the surface. For the molecules in monolayer 1 with dangling O atoms, i.e. both H atoms involved in hydrogen bonding, the remaining OH will find itself in an energetically unfavorable position, because of the



breaking of a hydrogen bond. It will therefore move away from its original position, in most cases 1-3 Å but sometimes also much longer. In monolayer 2, all molecules are four-coordinated, but the behavior is still very similar to the molecules with dangling O atoms in monolayer 1. For amorphous ice there exist no unique coordination sites as in crystalline ice. This explains the much less diverse behavior for the OH motion upon photoexcitation in different layers in amorphous ice.

The OH distance distribution for bilayer 2 is quite similar to that for bilayer 3 for both types of ice. Peak distances in bilayers 2 and 3 are around 1-2 Å for crystalline ice and about 1Å for amorphous ice. The average distances traveled by trapped OH for crystalline ice are 2.5 Å, 1.7 Å, and 1.7 Å for bilayers 1, 2, and 3, respectively, and for amorphous ice these numbers are 1.8 Å, 1.3 Å, and 1.2 Å. The shorter distances traveled in amorphous ice can in this case be attributed to the lack of shafts in the ice and/or the possibility that the formed OH radicals do not have to move as far as in crystalline ice to find an energetically more favorable position. The OH radicals in general do not move very far through the ice. A maximum of about 5 Å has been found for OH originating in monolayers 3 through 6 in crystalline ice and 4 through 6 in amorphous ice. The situation is however different for OH motion *over* the surface. For crystalline ice the maximum distance traveled over the surface is 85 Å with the maximum being 61 Å for amorphous ice. The strong OH-ice interaction is an important factor contributing to these long surface diffusion lengths. While a H atom moving over the surface is likely to desorb, the OH radical might not have sufficient kinetic energy to leave the surface but still have enough energy to sustain motion *parallel* to the surface. Because of the open structure of



the top monolayers of amorphous ice, OH formed in the third monolayer can move up to 23 Å, while the corresponding distance in crystalline ice is only 5 Å.

The vibrational state distributions of the trapped OH radicals are shown in Fig. 10, where a comparison is made between results for photoexcitation in the three bilayers with the corresponding gas-phase calculation. Since the vibrational energy in classical calculations is not quantized, a box quantization has been applied with vibrational energy levels taken from a Morse potential fitted to experiments.[58] The results are very similar for crystalline and amorphous ice and also comparing the three bilayers. The vibrational distributions are somewhat cooler in the ices compared to the gas phase, with an average vibrational energy of 0.5 eV, compared to 0.6 eV for the gas phase. This difference is of course explained by the intermolecular interactions the condensed-phase molecules experience. The differences between condensed phase and gas phase are small and the excitation energies are also different. Therefore it is difficult to make a general statement as to why less vibrationally excited OH radicals are formed in the ice than in the gas phase.

The vibrational state distribution in Fig. 10 corresponds to the vibrational energy immediatelly following dissociation, the energy being recorded once the dissociated OH bond becomes longer than 2 Å. Since OH is in a condensed-phase environment the vibrational energy is not a conserved quantity but the OH vibrational motion is coupled to the phonon modes of the ice. This coupling however appears to be weak since the average transferred energy per vibrational period is 3 meV in both crystalline and amorphous ice, which can be compared to the OH zero-point energy which is 0.23 eV.



The total average energy loss during the course of the trajectories is 9 meV over an average time of 0.8 ps in crystalline ice and 0.7 ps in amorphous ice. Part of the OH vibrational energy could possibly be transferred through near-resonant intermolecular vibrational energy if the surrounding $H_2O$ molecules were allowed to be flexible. This would potentially make the loss of vibrational energy much more efficient.

## G. Efficiency of $H_2O$ recombination

As seen in Fig. 4, there is a chance that the photofragments may recombine. Not all 'close encounters' of H and OH lead to recombination, however. One can analyze this by adopting two criteria for defining close encounters: (A) the intramolecular $H_2O$ energy, as given by the DK PES, is attractive by -0.1 eV, and (B) the intramolecular energy is -1.0 eV or less. In the first bilayer the fraction of trajectories obeying A is about 10% for both types of ice. Of these 50-60% go onto the stronger interaction case and of these finally 70-80% stay together as recombined $H_2O$. In bilayers 2 and 3 case A happens in 30-40% of the cases and 90% of these go on to case B. The recombination efficiency once the interaction is that strong is very close to unity (98%-100%). The lower recombination efficiency in bilayer 1 is due to the recombined $H_2O$ not interacting with the surroundings as strongly as deeper into the ice, making vibrational deexcitation slower. Redissociation of the molecule can also make either fragment desorb or travel a significant distance over the surface. In the 'cages' in the lower bilayers the fragments are much more likely to stay in close proximity of each other.



The reason for that the recombination probability is somewhat higher in amorphous ice than in crystalline ice (Section III.C) can be understood through the fact that there are larger fractions of H and OH that travel short distances in amorphous ice and do not escape the cage of the immediate surroundings compared to crystalline ice. Therefore the photofragments are more likely to be in close proximity of each other and are then also more likely to recombine.

## H. Recombined $H_2O$ distance distribution

The motion of the recombined $H_2O$ is defined analogously to the motion of OH. Fig. 11 displays the $H_2O$ distance distribution for bilayers 1 to 3 in both types of ice. These distributions are very similar to those for OH, the motion of OH being most important in determining where recombination takes place, because of its limited mobility compared to the H atom. Similarly to the OH radicals, motion of $H_2O$ through the bulk is restricted to a maximum of about 5 Å, while molecules recombined at the surface have been seen to travel 25 Å in crystalline ice and 32 Å in amorphous ice. The average distances for crystalline ice are 2.7 Å, 1.7 Å, and 1.7 Å for bilayers 1, 2, and 3, respectively, while the corresponding distances for amorphous ice are 3.0 Å, 1.6 Å, and 1.3 Å. The fact that these molecules can move significant distances at the surface and at least moderate distances in the bulk of the ice could have important implications for the possibility of phase transformation of ice under UV irradiation.[31]

## I.    Limitations of the present model



While we believe that the main trends in our results are essentially correct, there are a number of shortcomings that limit the accuracy of the model. The nature of the model can suggest how the results from these calculations may compare to experiments. The most important approximations in this work are: (i) the use of gas-phase PESs for the $H_2O$ intramolecular interactions, (ii) the use of (non-polarizable) pair potentials for the intermolecular interactions, (iii) freezing the intramolecular degrees of freedom of the surrounding $H_2O$ molecules, (iv) the simplified treatment of recombination, and (v) the use of classical dynamics for nuclear motion.

In Section II.B it was argued that the 'true' intermolecular potential for the first excited state of $H_2O$ should contain a long-range exchange-repulsion interaction. This would give a higher intermolecular energy than what is presently the case. Since the present potential reproduces well the experimental condensed-phase spectra this would mean that the intramolecular $H_2O$ excited-state interaction should be smaller, i.e. less repulsive. Therefore a different partitioning of the kinetic energy to intermolecular and intramolecular modes may be possible. A likely effect would be H atoms being released into the ice and into the gas phase with lower kinetic energies.

Also the $H$-$H_2O$ potential is likely somewhat too repulsive for condensed-phase simulations (see Section II.B). If this interaction were somewhat lower in energy and sustained the transient $H_3O$ species, the H atom may form a longer-lived complex with $H_2O$, thereby increasing the efficiency of kinetic energy transfer.



A full description of the processes in $H_2O$ ice photodissociation would require the surrounding $H_2O$ molecules to be fully flexible. This could lead to additional loss of H atom kinetic energy to intramolecular vibrational modes. Also the OH and recombined $H_2O$ vibrations could lose energy this way. Furthermore, the following reactions of H and $H_2O$ cannot be excluded at the high H atom kinetic energies involved:[25,46]

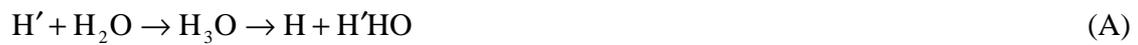

$$H' + H_2O \rightarrow H_3O \rightarrow H + H'HO \qquad\qquad (A)$$

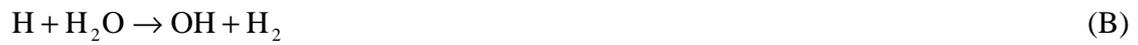

$$H + H_2O \rightarrow OH + H_2 \qquad\qquad (B)$$

If the hydrogen exchange reaction (A) occurs, it is likely that the released H atom will have lower kinetic energy than the initially reacting atom. If reaction B occurs there is a loss of H atoms, and $H_2$ is produced.

The above considerations taken together suggest that the calculations presented in this paper give an upper bound to the kinetic energy of the H atoms and a lower bound to the transfer of their kinetic energy to the ice. In turn, the computed distance over which the H atoms travel probably represents an upper bound and the kinetic energy distribution of the desorbing H atoms may be too hot. The desorption of surrounding $H_2O$ is probably overestimated because the initial H kinetic energy is likely somewhat high. As discussed in Section II.B the OH-$H_2O$ attractive potential is probably somewhat too weak. This suggests that the OH distances traveled may be overestimated, but it is not clear by how much.



Because we probably overestimate the distances traveled by the fragments, the probability of the photofragments coming together again is likely to be underestimated. However, in our model recombination only involves the ground state PES of $H_2O$. In reality, the outcome of this 'recombination' need not be reconstruction of $H_2O$ molecules. If H and OH come together on the first excited state, the complex will probably just redissociate into H and OH, while if this encounter takes place on a triplet PES, $O(^3P)$ + $H_2$ could be produced, analogously to the corresponding gas-phase process.[51] Another issue not explicitly covered in our model, is that even though the recombination would occur on the ground state PES, in the condensed phase the dissociation of a ground-state $H_2O$ molecule along a hydrogen bond would involve the transfer of a proton and not a hydrogen atom.[23] Therefore, if a $H_2O$ molecule is formed through recombination, because of its initially very high vibrational energy it could redissociate and, through transfer of a proton to a neighboring $H_2O$ molecule, form $H_3O^+$ and $OH^-$.

The importance of possible quantum effects is hard to estimate. As mentioned in Section II.C, classical gas-phase photodissociation calculations with purely repulsive excited states give good agreement with quantum dynamics. Regarding motion through the ice, the initially high kinetic energy of the H atom suggests that classical dynamics should be a reasonably good approximation. As the kinetic energy of H becomes smaller and its de Broglie wavelength larger, quantum effects will become more important. The statement above that the distances traveled by the H atoms are underestimated is actually only true as long as the motion of H is reasonably well described using classical dynamcs. At lower kinetic energies, tunneling through barriers to diffusion could become important. This



would then lead to *longer* distance traveled by the H atoms than predicted by a classical model. On much longer time scales than considered in our calculations, H atoms could potentially diffuse longer distances as well, either through tunneling or thermal hopping.

## IV.    SUMMARY AND OUTLOOK

A potential energy surface (PES) has been developed for describing the interaction of a $H_2O$ molecule in its first excited electronic state with surrounding $H_2O$ molecules in the condensed phase. The PES has been employed in a classical molecular dynamics model describing the photodissociation of a $H_2O$ molecule in the top six monolayers of crystalline and amorphous ice surfaces at 10 K. Overall the outcomes were found to be similar for the two types of ices, with a number of small differences in details.

The calculated absorption spectra of crystalline ice, amorphous ice and liquid water were found to be in good agreement with experiments. Upon photoexcitation in the top two (three) monolayers in crystalline (amorphous) ice the desorption of H atoms is the major outcome, while trapping of both H and OH and their recombination dominate in the lower layers.

The desorption of recombined $H_2O$ molecules was found to be rather small (0.2% per absorbed photon). Only $H_2O$ molecules formed from recombination of H and OH originating in the top two (three) monolayers desorbed. Indirect desorption, of $H_2O$ molecules not initially photodissociated in the top two monolayers, occurred in 0.2% of the trajectories for crystalline ice and in 0.04% of the ones for amorphous ice.



Desorption of H atoms is more probable from amorphous ice than for crystalline ice for all layers studied. The kinetic energy distribution of the desorbing H atoms is also hotter for amorphous ice than for crystalline ice. OH radicals formed from photodissociation can only desorb from the top two (three) monolayers in crystalline (amorphous) ice. The desorption probability of OH is much higher for crystalline ice than for amorphous ice.

H atoms on average move about 10 Å through the ice, but can actually move more than 60 Å before being trapped. OH radicals on average move only 1-2 Å and at most 5 Å through the ice, but OH released from photodissociation at the surface can move more than 80 Å over the surface. Recombined $H_2O$ molecules are found at an average distance of about 2 Å, and a maximum distance of 5 Å, from the $H_2O$ molecule initially photodissociated if dissociation occurs in the lower layers. If dissociation occurs at the surface, recombined $H_2O$ can move more than 30 Å over the surface, however.

The loss of H atom kinetic energy to phonon modes of the ice is more efficient in amorphous ice than in crystalline ice. This leads to slightly shorter average distances travelled in the ice for the H atoms that eventually become accomodated to the ice. However, the low density and open structure of the top monolayer in amorphous ice leads to fewer collisions for H atoms about to desorb, which results in the hotter H atom kinetic energy distribution mentioned above.



The findings for the mobility of the H atoms and OH radicals give clear indications about the possibility of these species to react with other atoms and molecules in the ice. H atoms, which clearly can move long distances through the ice, should be able to find a reaction partner within a radius of at least 60 Å from where it originated. OH radicals formed in the bulk of the ice would not be able to react with molecules which are not in their immediate vicinity (within a distance of 5 Å), while OH formed at the surface would be able to react with species at locations at least 80 Å from where it was formed.

Further developments of our model should involve improved descriptions of the potentials, incorporation of flexible $H_2O$ molecules, and a proper treatment of quantum effects. The study of possible further reactions of the photofragments with co-adsorbed species, such as CO, would also be very interesting, especially since this could help explain the composition of interstellar ices.

## ACKNOWLEDGEMENTS


We would like to thank Dr. R. van Harrevelt and Prof. M. C. van Hemert for providing us with a program for performing classical photodissociation of gas-phase $H_2O$. Dr. L. Valenzano is acknowledged for making a $H_2O$ dimer PES available to us. Some of the calculations reported here were carried out under a grant of computer time by the Dutch National Computing Facilities Foundation (NCF). This research was funded by a NWO Spinoza grant (for E. F. van Dishoeck) and a NWO-CW programme grant.

**Figure captions**

**Figure 1.** Top (top) and side view (bottom) of the six top monolayers in crystalline ice.

**Figure 2.** Top (top) and side view (bottom) of the six top monolayers in amorphous ice.

**Figure 3.** (a) Crystalline ice spectra calculated using the old and new potential model , calculated gas-phase $H_2O$ spectrum, and experimental absorption spectrum (first absorption band) (Ref. 10). (b) Absorption spectra for crystalline ice, amorphous ice, and liquid water calculated using the new potential model.

**Figure 4.** Probabilities of basic outcomes per monolayer in crystalline and amorphous ice using the new potential model.

**Figure 5.** Basic outcomes in the first and second monolayers of crystalline and amorphous ice using the new potential model. Note that the 'H desorbs' category is not included in the figure. 'OH desorbs' refers to the case when OH desorbs and H remains in the ice and 'HOH desorbs' is the case when the recombined $H_2O$ molecule desorbs.

**Figure 6.** Distibutions of the distances from the original positions of the trapped H atoms per bilayer in crystalline and amorphous ice using the new potential model.

**Figure 7.** Kinetic energy distributions of desorbing H atoms per bilayer from crystalline and amorphous ice using the new potential model.



**Figure 8.** Distributions of the distances from the original positions of the trapped OH radicals per bilayer for crystalline and amorphous ice using the new potential model.

**Figure 9.** Distributions of the distances from the original positions of the trapped OH radicals for the first two monolayers of crystalline and amorphous ice using the new potential model.

**Figure 10.** OH vibrational state distributions for gas-phase photodissociation and crystalline and amorphous ice photodissociation using the new potential model.

**Figure 11.** Distributions of the distances from the original positions of the trapped, recombined $H_2O$ molecules per bilayer for crystalline and amorphous ice using the new potential model.



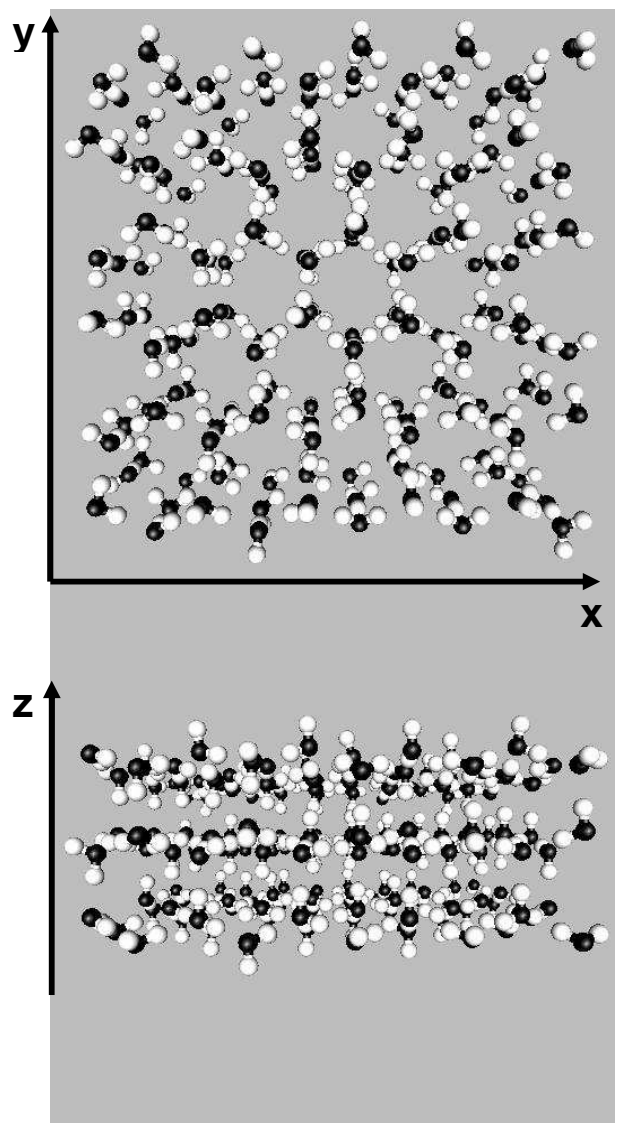

**Figure 1.** Andersson et al.



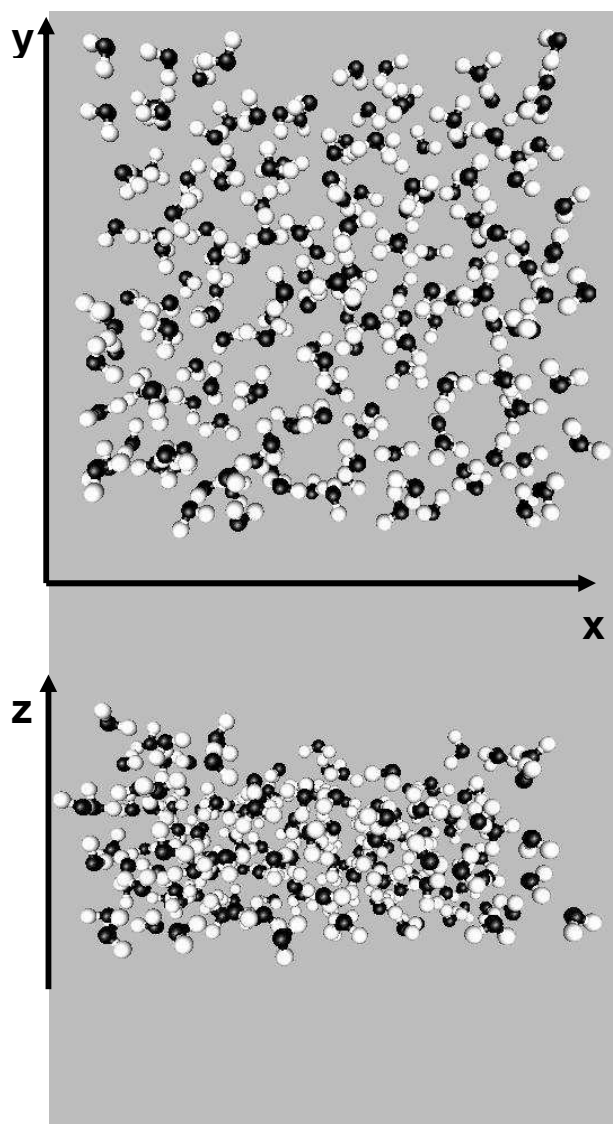

**Figure 2.** Andersson et al.



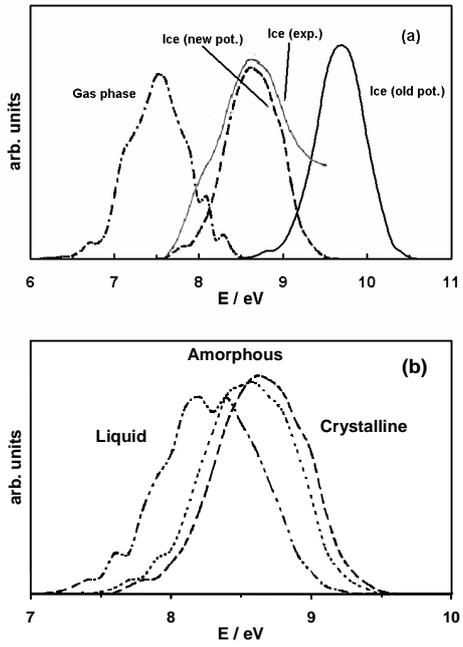

**Fig. 3** Andersson et al.



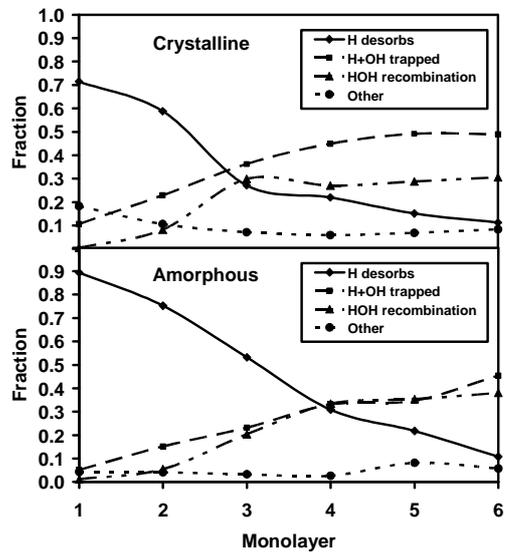

**Fig. 4** Andersson et al.



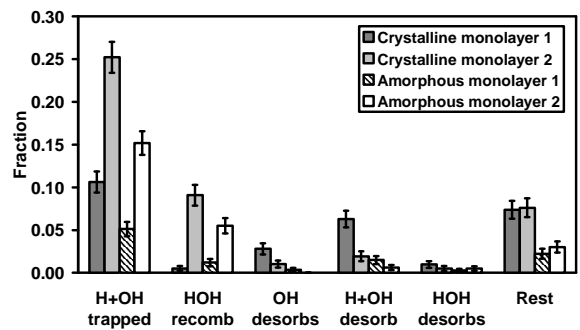

**Fig. 5** Andersson et al.



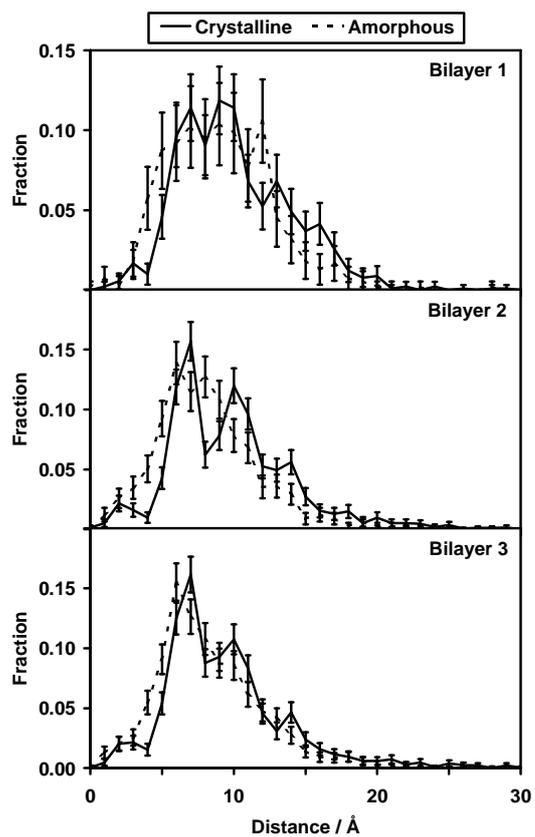

**Fig. 6** Andersson et al.



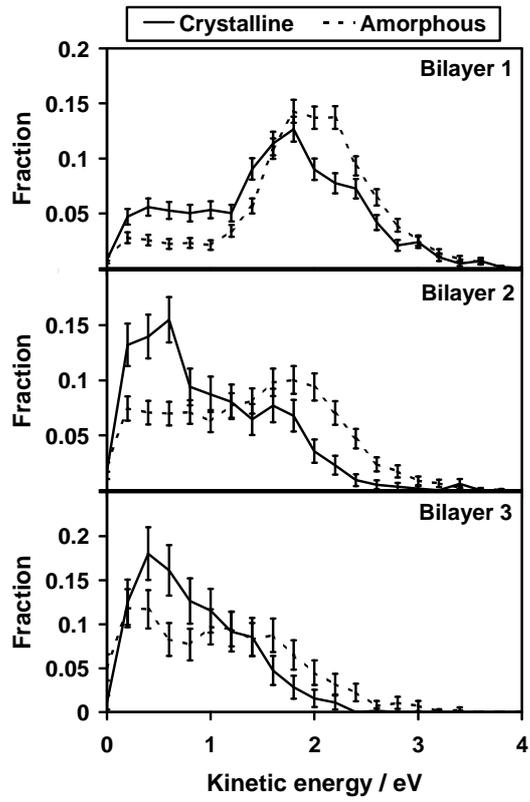

**Fig. 7** Andersson et al.



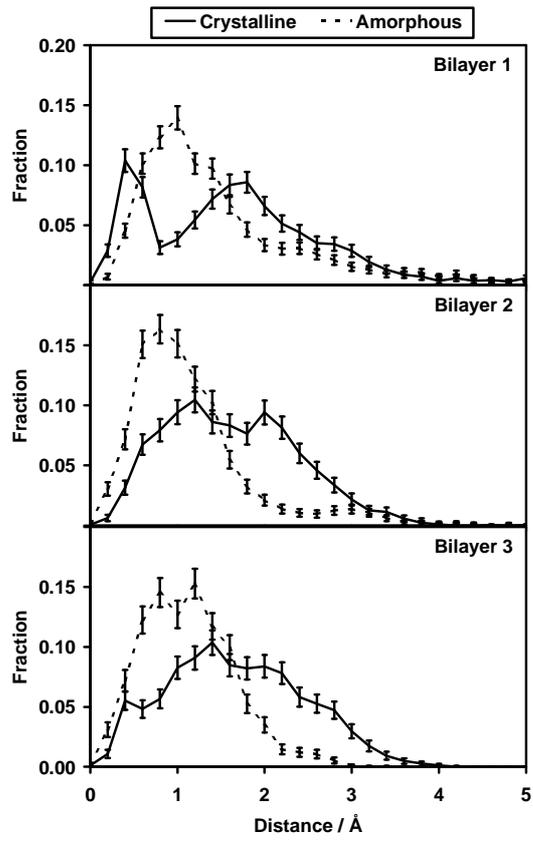

**Fig. 8** Andersson et al.



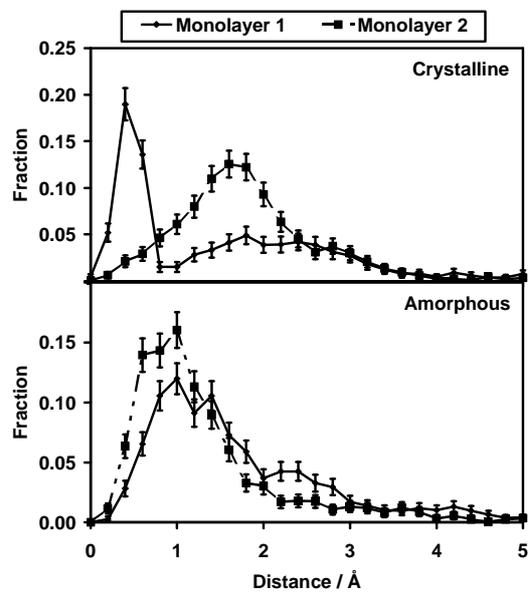

**Fig. 9** Andersson et al.



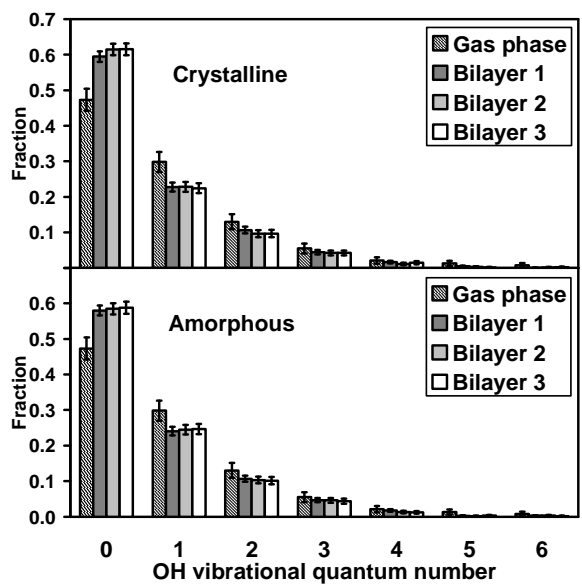

**Fig. 10** Andersson et al.



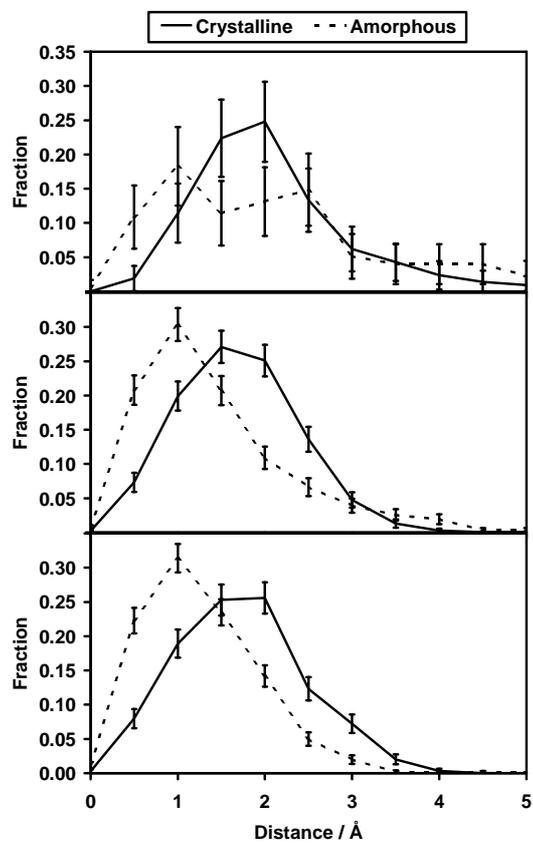

**Fig. 11** Andersson et al.



**Supporting material for 'Molecular dynamics study of photodissociation of water in crystalline and amorphous ice' by Andersson, Al-Halabi, Kroes, and van Dishoeck**

## THE H-H₂O POTENTIAL

The potential is given as follows (H is the free H atom, H(1) and H(2) are the H atoms in the $H_2O$ molecule, and '$i$' refers to the two H-H distances and the H-O distance):

$$V_{\text{H}-\text{H}_2\text{O}} = V_{disp}^{\text{H}}(R_{\text{HH}(1)}) + V_{disp}^{\text{H}}(R_{\text{HH}(2)}) + V_{disp}^{\text{H}}(R_{\text{HO}})$$
$$+ V_{rep}^{\text{H}}(R_{\text{HH}(1)}) + V_{rep}^{\text{H}}(R_{\text{HH}(2)}) + V_{rep}^{\text{H}}(R_{\text{HO}}) + V_{Morse}^{\text{H}}(R_{\text{HO}}) \tag{S1}$$

$$V_{disp}^{\text{H}}(R_i) = -D^{\text{H}}(R_i)C_6^i R_i^{-6} \tag{S2}$$

$$D^{\text{H}}(R_i) = \begin{cases} 1.0, & R_i \geq R_c^i \\ \exp\left[-\left(\dfrac{R_c^i}{R_i} - 1\right)^2\right], & R_i < R_c^i \end{cases} \tag{S3}$$

$$V_{rep}^{\text{H}}(R_i) = a_i \exp(-b_i R_i) \tag{S4}$$

$$V_{Morse}^{\text{H}}(R_{\text{OH}}) = D_{\text{OH}}\{\exp[-2\beta_{\text{OH}}(R_{e,\text{OH}} - R_{\text{OH}})] - 2\exp[-\beta_{\text{OH}}(R_{e,\text{OH}} - R_{\text{OH}})]\} \tag{S5}$$

The parameters are collected in Table I. All of these parameters have been obtained by refitting the YZCL2 PES.[46]

In the fitting procedure the $H_2O$ molecule was kept rigid in its TIP4P[42] geometry ($r_{\text{OH}}$ = 0.9572 Å; $\theta_{\text{HOH}}$ = 104.52°) and the H-$H_2O$ distance was varied for a number of different orientations. In Figs. S1 and S2 1D cuts are shown through our fitted potential, the



YZCL2 potential, and the H-H$_2$O potentials constructed by Bartels et al. (BHP)[61] and Zhang et al. (ZSB),[62] with the molecular orientations included in Fig. S1. For orientation A, the choice of the angle 117.3° of the H-O vector to the plane of the H$_2$O was made because this is the orientation of the H$_3$O local minimum on the YZCL2 PES. The BHP and ZSB potentials are also of a simple pair-potential form and, as far as we know, the only other published H-H$_2$O pair potentials. BHP was obtained in a 'semiempirical' fashion with the parameters being determined partly from theory and partly from experiment, while ZSB was first fitted to ab initio MP4 energy points (with basis set superposition error correction) and subsequently rescaled to reproduce a well depth derived from experiment. As can be seen in Fig. S1, all potentials agree reasonably well in the attractive 'low-energy' region, except in case D where the pair potentials give an interaction that is clearly more attractive than YZCL2. For the repulsive part of the potential shown in Fig. S2 it is seen that the present potential reproduces the YZCL2 PES surprisingly well, given its simple form. The ZSB potential is consistently too repulsive in this region and is actually only truly reliable in the attractive part of the potential. The BHP potential follows the YZCL2 PES rather well up to repulsive interactions of about 0.2-0.3 eV, which is surprising, since it was not fitted to any ab initio data. However, the present potential is clearly the best of these three pair potentials for the present simulations, in which H can emerge from the dissociated H$_2$O molecule with kinetic energies of a few eV.

**THE OH-H$_2$O POTENTIAL**

The interaction potential expression is given as:



$$V_{\text{OH}-\text{H}_2\text{O}} = V_{disp}^{\text{OH}}(R_{\text{OO}}) + V_{rep}^{\text{OH}}(R_{\text{OO}}) + V_{rep}^{\text{OH}}(R_{\text{OH}(1)}) + V_{rep}^{\text{OH}}(R_{\text{OH}(2)})$$
$$+ V_{rep}^{\text{OH}}(R_{\text{HO}(w)}) + V_{rep}^{\text{OH}}(R_{\text{HH}(1)}) + V_{rep}^{\text{OH}}(R_{\text{HH}(2)}) \tag{S6}$$
$$+ V_{elstat}^{\text{OH}}(R_{\text{OM}}, R_{\text{OH}(1)}, R_{\text{OH}(2)}, R_{\text{HM}}, R_{\text{HH}(1)}, R_{\text{HH}(2)})$$

$$V_{disp}^{\text{OH}}(R_{\text{OO}}) = -D^{\text{OH}}(R_{\text{OO}})C_6^{\text{OO}}R_{\text{OO}}^{-6} \tag{S7}$$

$$D^{\text{OH}}(R_{\text{OO}}) = \begin{cases} 1.0, & R_{\text{OO}} \geq 1.28 R_m^{\text{OO}} \\ \exp\left[-\left(\dfrac{1.28 R_m^{\text{OO}}}{R_{\text{OO}}} - 1\right)^2\right], & R_{\text{OO}} < 1.28 R_m^{\text{OO}} \end{cases} \tag{S8}$$

$$V_{elstat}^{\text{OH}} = \sum_{j=1}^{2} \sum_{k=1}^{3} q_j(R_{\text{OH}}) q_k R_{jk}^{-1} \tag{S9}$$

$$q_{\text{H}}(R_{\text{OH}}) = (a_q + b_q R_{\text{OH}}) \tag{S10}$$

$$q_{\text{O}}(R_{\text{OH}}) = -q_{\text{H}}(R_{\text{OH}}) \tag{S11}$$

The basic repulsion terms have the same form as in the H-$H_2O$ potential, the dispersion interaction is also treated the same way except for a slightly different $D(R)$ function (compare eqs. (S3) and (S8)). The TIP4P charges are used for the $H_2O$ molecule. In Eq. (S9), $q_j$ and $q_k$ are the charges on OH and $H_2O$, respectively. Also, $R_{\text{HO}(w)}$ is the distance between the H atom in OH and the O atom in $H_2O$, while $R_{\text{OH}}$ is the bond distance of OH. The parameters are collected in Table II. $R_m$ is calculated as the sum of two oxygen van der Waals radii[63] plus 0.1 Å analogous to the $R_m$ for Cl-O dispersion in Ref. 64.

A comparison between the fitted OH-$H_2O$ potential and a corresponding *ab initio* potential is presented in Figs. S3 and S4. Here six 1D cuts through the fitted potential are presented alongside potential curves calculated with CCSD(T). These coupled-cluster



calculations were performed with an aug-cc-pVDZ basis set[65] using the Gaussian 03 program package.[66] For these six orientations it is seen that the simple pair potential agrees surprisingly well with the *ab initio* potential, given the fact that the pair potential was not fitted to these *ab initio* energies. The attractive interaction is however almost everywhere somewhat smaller in the pair potential. The fraction of the pair potential to the CCSD(T) potential is about 0.8 at the minima along the contours for structures I, II, and IV, about 0.5 for II and VI, but only 0.25 for structure V. At longer range agreement is better and at an O-H distance of 3.0 Å the fraction is 0.9-1.0 for I, II, and IV, about 0.6 for II and VI, and 0.4 for structure V.

## SWITCHING FUNCTIONS

To smoothly connect the different parts of the potentials, i.e. between where the dissociating $H_2O$ is still intact to where it is dissociated into H and OH fragments, a number of switching functions have been devised. Beginning with the electrostatic potential the switch is made for the charges according to:

$$q_O = f(R_{OH_a})f(R_{OH_b})q_O^{TIP3P} + (1 - f(R_{OH_a})f(R_{OH_b})) \times \\ (q_O^{OH}(R_{OH_a}) + q_O^{OH}(R_{OH_b}))$$

(S12)

$$q_{H_n} = f(R_{OH_a})f(R_{OH_b})q_H^{TIP3P} + (1 - f(R_{OH_a})f(R_{OH_b})) \times \\ q_H^{OH}(R_{OH_n})$$

(S13)

$$q_O^{OH}(R_{OH_n}) = -g(R_{OH_n})(a_q + b_q R_{OH_n})$$

(S14)

$$q_H^{OH}(R_{OH_n}) = -q_O^{OH}(R_{OH_n})$$

(S15)



$$f(R) = \begin{cases} 0.0, & R \geq R_2 \\ 2(1-x)^3, & \dfrac{R_1 + R_2}{2} \leq R < R_2 \\ 1 - 6x^2 + 6x^3, & R_1 < R < \dfrac{R_1 + R_2}{2} \\ 1.0, & R \leq R_1 \end{cases} \tag{S16}$$

$$x = \frac{R - R_1}{R_2 - R_1} \tag{S17}$$

$$g(R) = f(R) \tag{S18}$$

Here, $OH_a$ and $OH_b$ refer to the O-H distances within the dissociating $H_2O$ molecule with $OH_n$ being either $OH_a$ or $OH_b$. The form of the switching function is the same as used for long-range cutoff in Ref. 64, and the cutoff distances for the switch are $R_1 = 1.1$ Å and $R_2 = 1.6$ Å. These distances were chosen on the basis of comparing the present potential to 3D representations of the ground- and first-excited state $H_2O$-$H_2O$ PESs[67,68] and thereby seeing where a switch should be reasonable to make if neighboring $H_2O$ molecules are taken into account. Note that even though the $f$ and $g$ switching functions are taken to be equal here this is not a necessary condition. In this case, however, the switches have the same functional form and share the same $R_1$ and $R_2$ parameters. This switching scheme guarantees that the system stays charge neutral at all times. Note that the OH charges will be completely damped out at OH distances of 1.6 Å and larger in contrast to the original OH-$H_2O$ potential.

For the dispersion interactions similar switches apply:



$$V_{disp}^{O} = -f(R_{OH_a})f(R_{OH_b})\frac{C^{TIP3P}}{R_{OO}^{6}} + (1 - f(R_{OH_a})f(R_{OH_b})) \times$$

$$V_{disp}^{OH}(R_{OO}) \tag{S19}$$

$$V_{disp}^{H_n} = h(R_{H_nO})\left[V_{disp}^{H}(R_{H_nO_e}) + V_{disp}^{H}(R_{H_nH(1)}) + V_{disp}^{H}(R_{H_nH(2)})\right] \tag{S20}$$

$$h(R) = \begin{cases} 1.0, & R \geq R_2 \\ 1 - 6y^2 + 6y^3, & \dfrac{R_1 + R_2}{2} \leq R < R_2 \\ 2(1-y)^3, & R_1 < R < \dfrac{R_1 + R_2}{2} \\ 0.0, & R \leq R_1 \end{cases} \tag{S21}$$

Here $O_e$ refers to an *external* O atom, i.e. belonging to one of the surrounding $H_2O$ molecules, and $H(1)$ and $H(2)$ as above also belong to the surrounding molecules. $R_1$ and $R_2$ have the same values as for the *f* and *g* switches. The O-O dispersion interaction is switched between the TIP3P and the OH-$H_2O$ potential, which in this case happens to be the same. When the H atom is formed during dissociation, the H-$H_2O$ dispersion potential is switched on through the use of the *h* function.

The repulsive interactions are switched in a completely analogous way to the dispersion:

$$V_{rep}^{O} = f(R_{OH_a})f(R_{OH_b})\frac{A^{TIP3P}}{R_{OO}^{12}} + (1 - f(R_{OH_a})f(R_{OH_b})) \times$$

$$\left[V_{rep}^{OH}(R_{OO}) + V_{rep}^{OH}(R_{OH(1)}) + V_{rep}^{OH}(R_{OH(2)})\right] \tag{S22}$$

$$V_{rep}^{H_n} = (1 - h(R_{H_nO}))\left[V_{rep}^{OH}(R_{H_nO_e}) + V_{rep}^{OH}(R_{H_nH(1)}) + V_{rep}^{OH}(R_{H_nH(2)})\right]$$

$$+ h(R_{H_nO})\left[V_{rep}^{H}(R_{H_nO_e}) + V_{rep}^{H}(R_{H_nH(1)}) + V_{rep}^{H}(R_{H_nH(2)})\right] \tag{S23}$$



The Morse potential is, just like the other H-H$_2$O interactions, switched on through the $h$ switch:

$$V_{Morse}^{H_n} = h(R_{H_nO})V_{Morse}^{H}(R_{H_nO_e})$$  (S24)

Note that in the limits of O+H+H and O+H$_2$ fragments, the O atom will interact with H$_2$O through the dispersion and repulsive interactions of O in OH, and the hydrogen atoms (also in H$_2$) will interact through the H-H$_2$O potential. This is of course not optimal for H$_2$, but in our case H$_2$ is rarely formed (sse Section III.C).

For the dissociating H$_2$O molecule the intramolecular PES is initially the first excited state PES. When the molecule is dissociated and the first excited state has become near degenerate to the ground state (both states correlate asymptotically to H($^2$S) + OH(X$^2\Pi$)), a smooth switch is made to the ground state PES. This is made on the assumption that an internal conversion will be efficient through large mixing of the electronic states, possibly induced by interaction with the surrounding molecules. The potential in the switching region is written as a linear combination of the ground and excited-state PESs:

$$V_{H_2O} = s(R_{OH_a})s(R_{OH_b})V_{H_2O}^{es} + (1 - s(R_{OH_a})s(R_{OH_b}))V_{H_2O}^{gs}$$  (S25)

The $s$ switching function has the same functional form as the $f$ function (Eqs. (S16) and (S17)) with the difference that $R_1 = 3.0$ Å and $R_2 = 3.5$ Å. These distances have been chosen because this is where the excited-state and ground-state PESs become near degenerate and the switch can be made without introducing troublesome kinks in the potential, the difference in energy between the PESs being smaller than 0.05 eV. If the



dissociation of the OH bond should be reversed before reaching 3.5 Å, the system will go back to the excited state PES if the bond becomes shorter than 3 Å. When the OH bond becomes longer than 3.5 Å the system will remain on the ground state PES. This allows for recombination of H and OH.

**Table I.** Parameters for the H-$H_2O$ potential.

Dispersion

| | |
|---|---|
| $C_6^{HH}$ | 30.2562 kJ mol$^{-1}$ Å$^6$ |
| $R_c^{HH}$ | 2.81541 Å |
| $C_6^{HO}$ | 2216.76 kJ mol$^{-1}$ Å$^6$ |
| $R_c^{HO}$ | 5.18068 Å |

Repulsion

| | |
|---|---|
| $a_{HH}^{H}$ | 5563.57 kJ mol$^{-1}$ |
| $b_{HH}^{H}$ | 4.27352 Å$^{-1}$ |
| $a_{HO}^{H}$ | 4158.88 kJ mol$^{-1}$ |
| $b_{HO}^{H}$ | 2.63805 Å$^{-1}$ |

Morse

| | |
|---|---|
| $D_{OH}$ | 287.598 kJ mol$^{-1}$ |
| $\beta_{OH}$ | 4.73771 Å$^{-1}$ |
| $R_{e,OH}$ | 0.816286 Å |



**Table II.** Parameters for the OH-H$_2$O potential.

| | |
|---|---|
| Dispersion | |
| $C_6^{OO}$ | 2489.48 kJ mol$^{-1}$ Å$^6$ |
| $R_m^{OO}$ | 3.14 Å |
| | |
| Repulsion | |
| $a_{OO}^{OH}$ | 299296 kJ mol$^{-1}$ |
| $b_{OO}^{OH}$ | 3.970 Å$^{-1}$ |
| $a_{OH}^{OH} = a_{HO}^{OH}$ | 17088 kJ mol$^{-1}$ |
| $b_{OH}^{OH} = b_{HO}^{OH}$ | 3.914 Å$^{-1}$ |
| $a_{HH}^{OH}$ | 3263 kJ mol$^{-1}$ |
| $b_{HH}^{OH}$ | 3.127 Å$^{-1}$ |
| | |
| Electrostatic | |
| $a_q$ | 0.69004 $e$ |
| $b_q$ | -0.30648 $e$ Å$^{-1}$ |



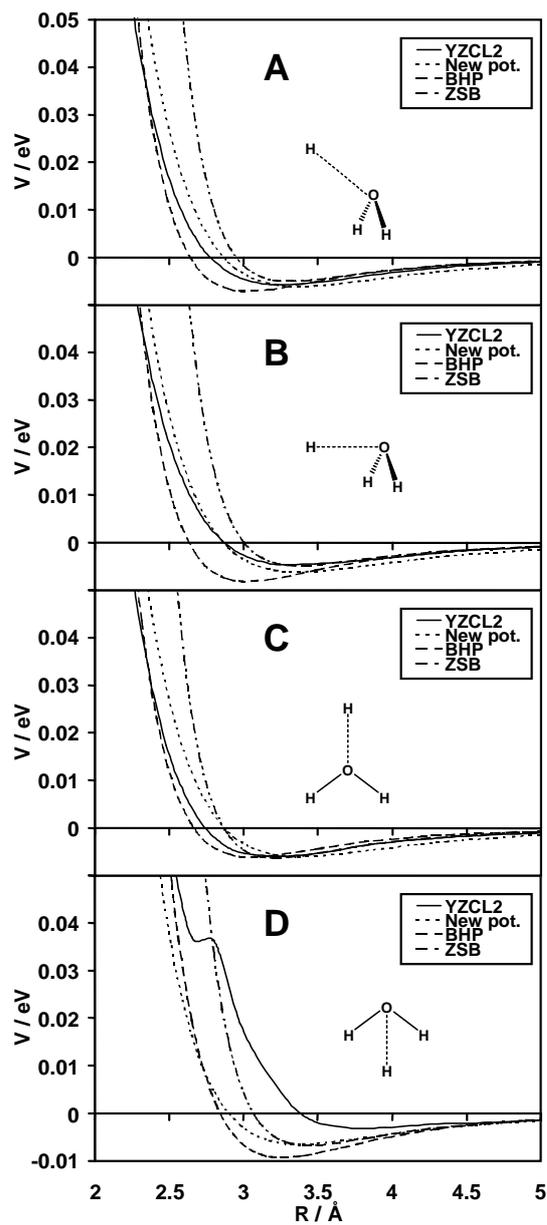

**Figure S1.** Low-energy part of the H-H$_2$O interaction potential as a function of H-O distance for four different orientations (see inserts).



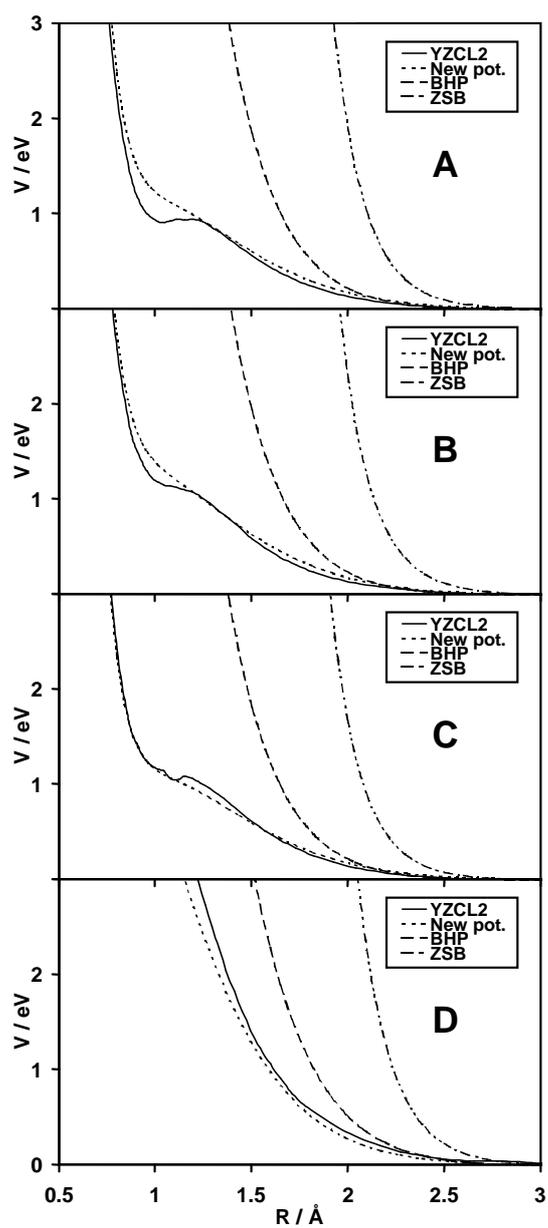

**Figure S2.** High-energy part of the H-H$_2$O interaction potential as a function of H-O distance for four different orientations (see inserts in Fig. 3).



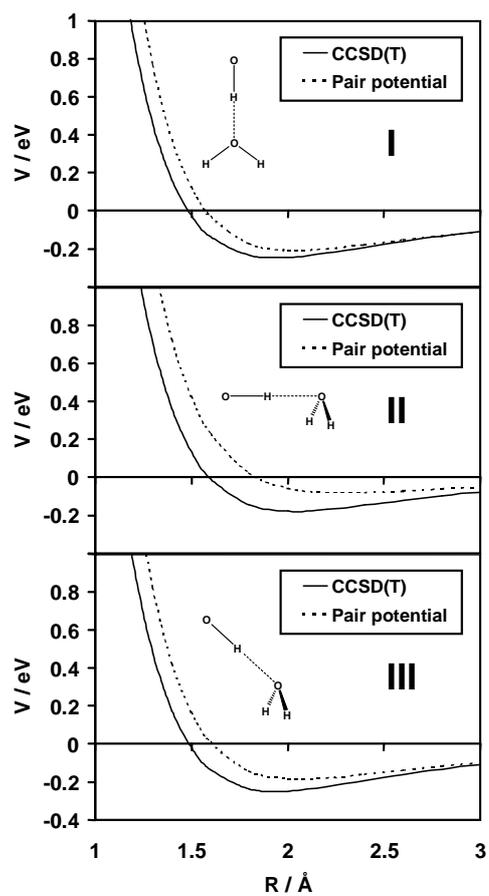

**Figure S3.** 1D cuts through the OH-H$_2$O potential energy surface given by CCSD(T) and the presently used pair potential for three different orientations (see inserts).



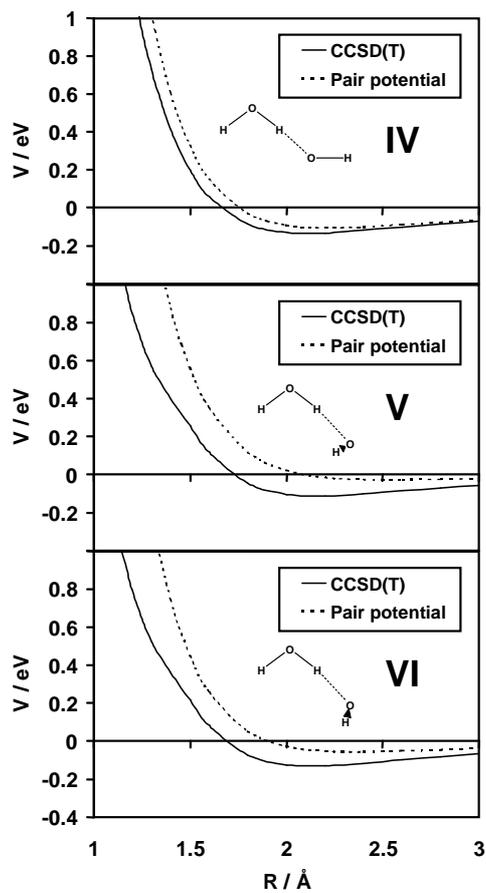

**Figure S4.** 1D cuts through the OH-H$_2$O potential energy surface given by CCSD(T) and the presently used pair potential for three different orientations (see inserts).



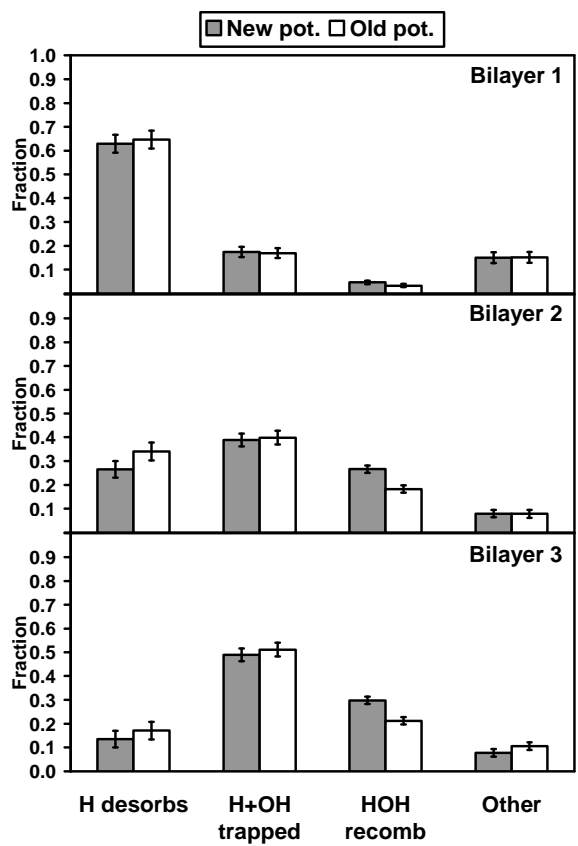

**Figure S5.** Probabilities of basic outcomes of photodissociation calculations per bilayer using the old and new potential models.



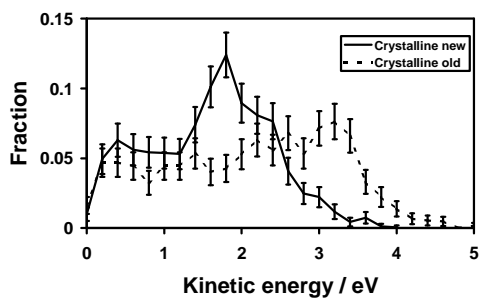

**Figure S6.** Kinetic energy distributions of desorbing H atoms from photodissociation using the old and new potential models.